\documentclass[journal]{IEEEtran}
\usepackage{cite,amsmath,amssymb,pifont}
\usepackage{color}
\usepackage{graphicx}
\usepackage{pstool}
\usepackage{overpic}
\usepackage[usestackEOL]{stackengine}
\usepackage{array,booktabs,multirow}

\usepackage{courier}
\usepackage{listings}
\usepackage{dblfloatfix} 

\usepackage[subrefformat=parens,labelformat=parens]{subfig}

\newcommand{\ve}[1]{\mathbf{#1}}
\newcommand{\te}[1]{\overline{\overline{#1}}}

\newcolumntype{N}{@{}m{0pt}@{}}

\newcounter{tempEquationCounter}
\newcounter{thisEquationNumber}
\newenvironment{floatEq}
{\setcounter{thisEquationNumber}{\value{equation}}\addtocounter{equation}{1}
\begin{figure*}[!b]
	\hrulefill\vspace*{4pt}
\normalsize\setcounter{tempEquationCounter}{\value{equation}}
\setcounter{equation}{\value{thisEquationNumber}}
}
{\setcounter{equation}{\value{tempEquationCounter}}
\end{figure*}
}

\newcommand{\htext}[1]{%
	\makebox[0pt]{\Centerstack{#1}}
}

\newcommand{\vtext}[1]{%
	\makebox[0pt]{\rotatebox[origin=c]{90}{\Centerstack{#1}}}
}

\begin{document}

\title{Multipolar Modeling of \\Spatially Dispersive Metasurfaces}

\author{Karim~Achouri, Ville Tiukuvaara, and Olivier J. F. Martin}

\maketitle

\begin{abstract}
There is today a growing need to accurately model the angular scattering response of metasurfaces for optical analog processing applications. However, the current metasurface modeling techniques are not well suited for such a task since they are limited to small angular spectrum transformations, as shall be demonstrated. The goal of this work is to overcome this limitation by improving the modeling accuracy of these techniques and, specifically, to provide a better description of the angular response of metasurfaces. This is achieved by extending the current methods, which are restricted to dipolar responses and weak spatially dispersive effects, so as to include quadrupolar responses and higher-order spatially dispersive components, for a metasurface in a uniform environment. The accuracy of the newly derived multipolar model is demonstrated by predicting the angular scattering of a dielectric metasurface placed in a vacuum. This results in a modeling accuracy that is at least 3.5 times better than the standard dipolar model. 
\end{abstract}	


\IEEEpeerreviewmaketitle


\section{Introduction}

Over the past decade, the research field of metasurfaces -- the two-dimensional counterparts of three-dimensional metamaterials -- has spectacularly flourished and has led to a plethora of metasurface concepts and applications~\cite{landy2008perfect,schurig2006metamaterial,kildishev2013planar,yu2014flat,chenHuygensMetasurfacesMicrowaves2018,achouri2020electromagnetic}. To help designing metasurfaces, one of the most common modeling approach has been to treat them as discontinuous electromagnetic sheets harboring effective dipolar responses~\cite{6477089,achouri2014general,epstein2016arbitrary}. Over the past few years, these metasurface modeling techniques have shown to be particularly effective at describing the complex electromagnetic responses of these structures and have enabled the implementation of optimized transformations such as perfect anomalous reflection and refraction~\cite{diazrubioGeneralizedReflectionLaw2017,Lavigne2018}.

Today, there is a growing interest towards the design of metasurface-based spatial processors for optical analog signal processing, as evidenced by the various examples of spatial differentiators and integrators that have already been  reported in the literature~\cite{porsAnalogComputingUsing2015,chizariAnalogOpticalComputing2016,abdollahramezaniDielectricMetasurfacesSolve2017,babashahIntegrationAnalogOptical2017,zhuPlasmonicComputingSpatial2017,momeni2019generalized,zhangControllingAngularDispersions2020}. From these studies, it is apparent that the successful implementation of more advanced signal processing operations clearly requires a very accurate control of the angular scattering response of a metasurface. Naturally, the metasurface modeling techniques mentioned above may be used as powerful design tools for realizing such operations efficiently.  

However, a major drawback of these techniques is that, although they work well in the paraxial regime or for illuminations with a fixed direction of propagation~\cite{angularAchouri2020}, they are currently rather limited when it comes to properly modeling metasurface operations requiring large angular spectrum, as we shall demonstrate thereafter.

In order to overcome this limitation, we propose to improve the angular modeling accuracy of these techniques by deriving a model that includes higher-order multipolar moments and their associated spatially dispersive components~\cite{raabMultipoleTheoryElectromagnetism2005,agranovich2013crystal,simovski2018composite}. Indeed, the main limiting factor of the current models is that they only take into account dipolar responses and only incorporate weak spatially dispersive effects. This is unfortunately not well suited for designing operations that deal with large angular spectrum since it is precisely in such cases that spatial dispersion starts to play a significant role~\cite{choContributionElectricQuadrupole2008,petschulatUnderstandingElectricMagnetic2010,muhligMultipoleAnalysisMetaatoms2011,grahnElectromagneticMultipoleTheory2012,evlyukhinCollectiveResonancesMetal2012,yaghjianHomogenizationSpatiallyDispersive2013,silveirinhaBoundaryConditionsQuadrupolar2014,Koendrink2014Retrieval,poutrina2014multipole,babichevaMetasurfacesElectricQuadrupole2018}. Therefore, our goal is to derive a model that includes both dipolar and quadrupolar moments and all the associated hypersusceptibility components, many of which are related to spatial dispersion~\cite{achouri2021extension}. While focusing on this goal, we will assume the metasurface to have the same media on either side.

For self-consistency, we first review the conventional metasurface modeling approach based on electric and magnetic dipoles in Sec.~\ref{sec:GenCon}. Then, in Sec.~\ref{sec:lim}, we show its limitations when it comes to modeling the angular scattering response of a metasurface. The extended model that includes both dipolar and quadrupolar moments is then derived in Sec.~\ref{sec:Deriv} and the associated concept of spatial dispersion is discussed in Sec.~\ref{sec:SD}. An example is then presented in Sec.~\ref{sec:example} to illustrate the method and demonstrate its performance. Finally, we conclude in Sec.~\ref{sec:concl}.

\section{Dipolar Modeling of Metasurfaces}

\subsection{General Concepts}
\label{sec:GenCon}

A common method for modeling an electromagnetic metasurface is to assume that it may be reduced to a discontinuous interface consisting of a zero-thickness fictitious sheet supporting electric and magnetic polarization currents~\cite{holloway2012overview,6477089,Pfeiffer2013a,PhysRevApplied.2.044011,6905746,epstein2016arbitrary,achouri2014general,Achouri2015c,achouri2018design}. The interactions of the metasurface with an exciting electromagnetic field may then be modeled using appropriate boundary conditions -- the Generalized Sheet Transition Conditions (GSTCs) -- that relate incident and scattered fields to the dipolar moments induced on the metasurface~\cite{Idemen1973,kuester2003av}. For a metasurface lying in the $xy$-plane at $z=0$, the frequency-domain\footnote{The time dependence $e^{j\omega t}$ is assumed and omitted throughout.} GSTCs read
\begin{subequations}
	\label{eq:GSTCsTD}
	\begin{align}
        \hat{\ve{z}}\times\Delta\ve{H}&=+j\omega\ve{P}_\parallel - \hat{\ve{z}}\times \nabla M_z,\\
		\hat{\ve{z}}\times\Delta\ve{E}&=-j\omega\mu_0\ve{M}_\parallel - \frac{1}{\epsilon_0}\hat{\ve{z}}\times \nabla P_z,
	\end{align}
\end{subequations}
where $\Delta\ve{H} = \ve{H}^+ - \ve{H}^-$ and $\Delta\ve{E} = \ve{E}^+- \ve{E}^-$ are respectively the differences of the fields between both sides of the metasurface with `$+$' and `$-$' indicating the fields at $z=0^+$ and $z=0^-$, and $\ve{P}_\parallel$ and $\ve{M}_\parallel$ are the \textit{tangential} electric and magnetic polarization densities induced on the metasurface, respectively. Under such a dipolar approximation, $\ve{P}$ and $\ve{M}$ may generally be expressed in terms of the metasurface effective bianisotropic surface susceptibility tensors $\te{\chi}_\text{ee}$, $\te{\chi}_\text{mm}$, $\te{\chi}_\text{em}$ and $\te{\chi}_\text{me}$ as~\cite{6477089,PhysRevApplied.2.044011,epstein2016arbitrary,achouri2014general,Achouri2015c,achouri2018design}
\begin{equation}\label{eq:PM}
	\begin{bmatrix}
		\ve{P}\\
		\ve{M}
	\end{bmatrix}=
\begin{bmatrix}
	\epsilon_0\te{\chi}_\text{ee} & \frac{1}{c_0}\te{\chi}_\text{em}\\
	\frac{1}{\eta_0}\te{\chi}_\text{me} & \te{\chi}_\text{mm}
\end{bmatrix}
\cdot
\begin{bmatrix}
	\ve{E}_\text{av}\\
	\ve{H}_\text{av}
\end{bmatrix},
\end{equation}
where $\ve{E}_\text{av} = (\ve{E}^++\ve{E}^-)/2$ and $\ve{H}_\text{av} = (\ve{H}^++\ve{H}^-)/2$ are the average electric and magnetic fields on the metasurface, respectively.

Combining~\eqref{eq:PM} with~\eqref{eq:GSTCsTD} leads to a system of equations that may either be used to predict the scattering response of a metasurface with known susceptibilities or to compute the metasurface susceptibilities required to achieve a desired scattering behavior specified in terms of incident and scattered fields. While these equations have proven to be very effective at modeling and synthesizing the vast majority of metasurfaces, they still suffer from an important limitation, which is their limited capability to accurately model the \textit{angular scattering response} of a metasurface~\cite{angularAchouri2020}.
%
%

\subsection{Limitations of the Model}
\label{sec:lim}

To illustrate the limitations of~\eqref{eq:GSTCsTD}, we consider the following simple example: model the angular scattering response of a uniform metasurface composed of a subwavelength periodic array of lossless dielectric cylinders. The metasurface is illuminated by a TM-polarized plane wave propagating in the $+z$ direction, as illustrated in Fig.~\ref{fig:MS}, where the plane of incidence is limited to the $xz$-plane, for simplicity. The media on both sides of the metasurface are vacuum. 
\begin{figure}[h!]
	\centering
	\includegraphics[width=0.7\columnwidth]{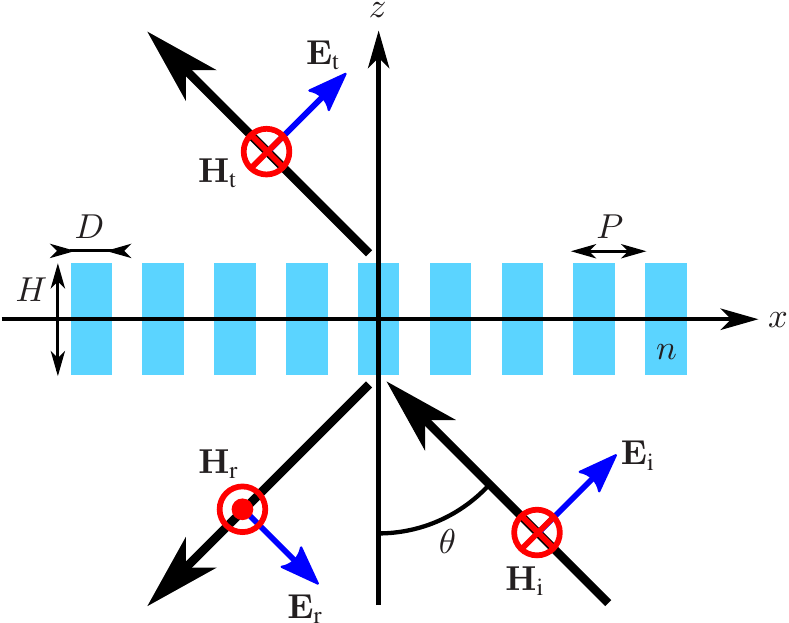}
	\caption{Cross-section of a uniform metasurface composed of a periodic array of dielectric cylinders illuminated by an obliquely propagating TM-polarized plane wave.}
	\label{fig:MS}
\end{figure}
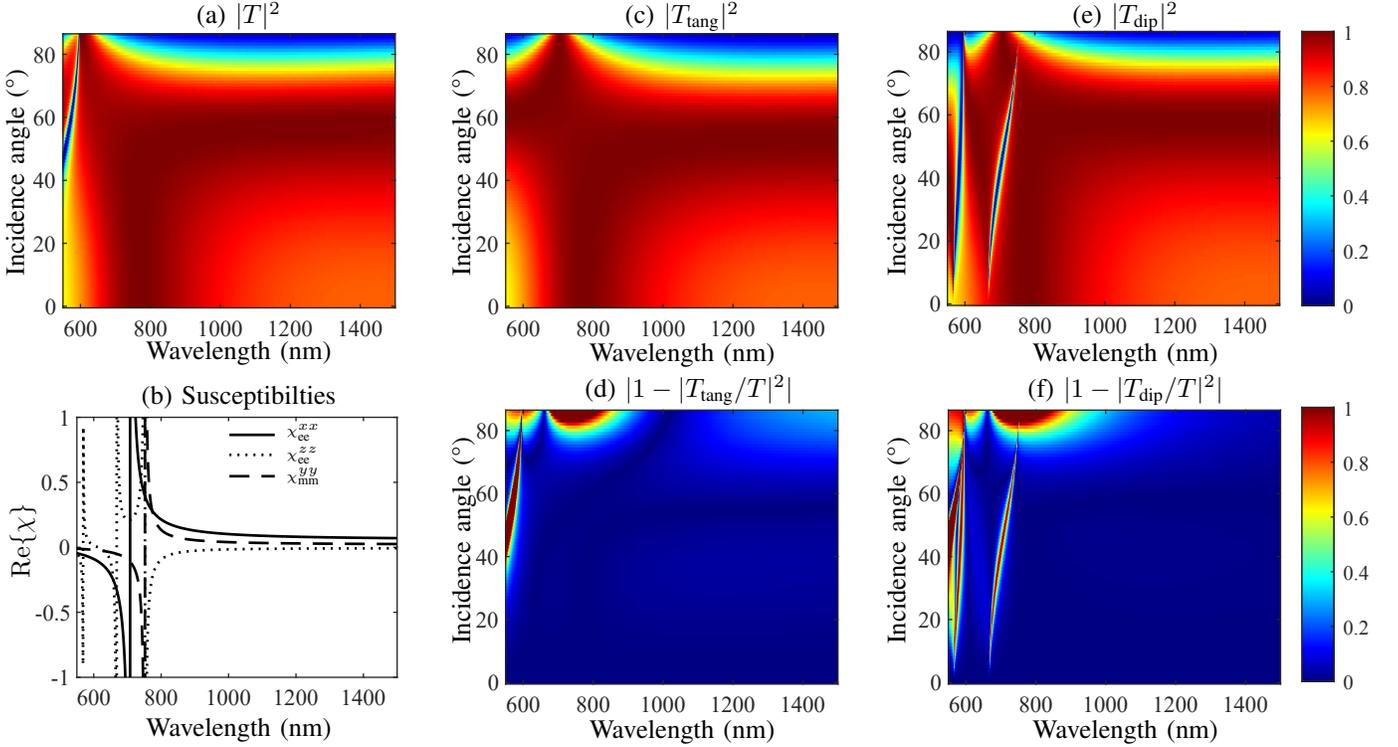
\begin{figure*}[t!]
	\centering
	\begin{overpic}[width=\textwidth,grid=false,trim={0cm 0cm 0cm 0cm},clip]{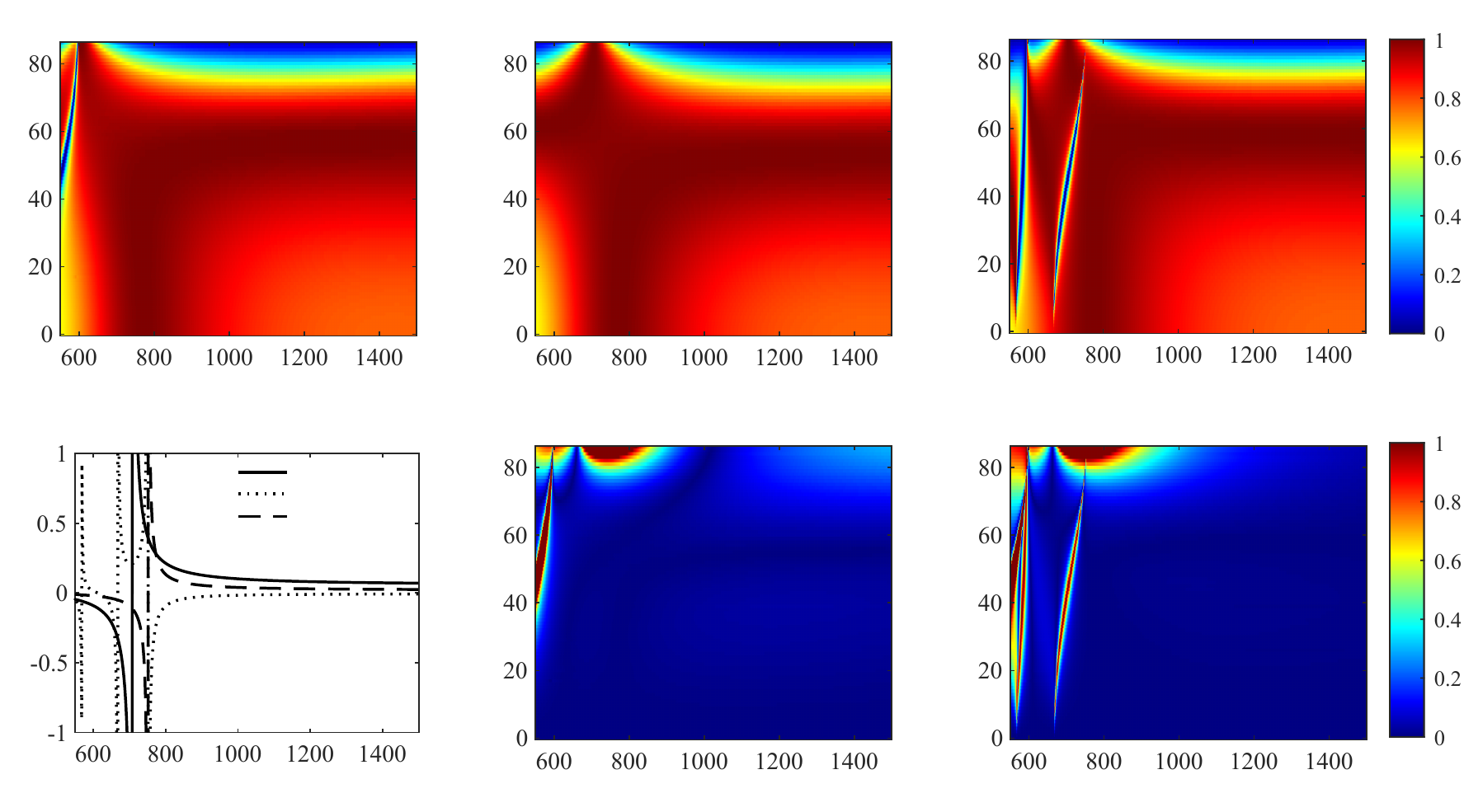}
		\put(17,1.1){\htext{Wavelength (nm)}}
		\put(49.5,1){\htext{Wavelength (nm)}}
		\put(82,1){\htext{Wavelength (nm)}}
		\put(17,28.8){\htext{Wavelength (nm)}}
		\put(49.5,28.6){\htext{Wavelength (nm)}}
		\put(82,28.6){\htext{Wavelength (nm)}}
		\put(1.2,15){\vtext{$\operatorname{Re}\{\chi\}$}}
		\put(33.5,15){\vtext{Incidence angle (°)}}
		\put(65.4,15){\vtext{Incidence angle (°)}}
		\put(0.8,42){\vtext{Incidence angle (°)}}
		\put(33.5,42){\vtext{Incidence angle (°)}}
		\put(65.4,42){\vtext{Incidence angle (°)}}
		\put(17,25.5){\htext{(b) Susceptibilties}}
		\put(50,26){\htext{(d) $|1-|T_\text{tang}/T|^2|$}}
		\put(82,26){\htext{(f) $|1-|T_\text{dip}/T|^2|$}}
		\put(17,53.4){\htext{(a) $|T|^2$}}
		\put(49.5,53.4){\htext{(c) $|T_\text{tang}|^2$}}
		\put(82,53.4){\htext{(e) $|T_\text{dip}|^2$}}
		\put(20.5,23.2){\tiny{}$\chi_\text{ee}^{xx}$}
		\put(20.5,21.5){\tiny{}$\chi_\text{ee}^{zz}$}
		\put(20.5,19.9){\tiny{}$\chi_\text{mm}^{yy}$}
	\end{overpic}
	\caption{Response of the metasurface in Fig.~\ref{fig:MS} with $P=225$~nm, $D=200$~nm, $H=200$~nm and $n=2.55$.  (a)~Full-wave simulated transmission coefficient, $|T|^2$. (b)~Retrieved real part of the metasurface tangential electric (solid line), normal electric (dotted line), and tangential magnetic (dashed line) susceptibilities using~\eqref{eq:Xret}. (c)~Predicted transmission coefficient using the tangential dipolar approximation, using~\eqref{eq:TRpred} with $\chi_\text{ee}^{zz}=0$. (d)~Relative transmission error. (e-f) Predicted transmission coefficient and relative error with complete dipolar susceptibilities using \eqref{eq:TRpred}. }
	\label{fig:Limit}
\end{figure*}

To model the interaction depicted in Fig.~\ref{fig:MS}, we define the fields that take part in it as
\begin{subequations}\label{eq:fields}
	\begin{align}
			E_{x,\text{a}} &= +A_\text{a}\frac{k_{z,\text{a}}}{k}e^{\mp j(k_xx + k_{z,\text{a}}z)},\\
			E_{z,\text{a}} &= -A_\text{a}\frac{k_x}{k}e^{\mp j(k_xx + k_{z,\text{a}}z)},\\
			H_{y,\text{a}} &= \pm\frac{A_\text{a}}{\eta_0}e^{\mp j(k_xx + k_{z,\text{a}}z)},
	\end{align}
\end{subequations}
where $\text{a}=\{\text{i},\text{r},\text{t}\}$ to differentiate between the incident, reflected and transmitted waves, $A_\text{a} = \{1,-R,T\}$ with $R$ and $T$ being the complex reflection and transmission coefficients, respectively, and $k_{z,\text{a}}=\{k_z,-k_z,k_z\}$. Note that since the metasurface period is subwavelength, all waves share the same tangential wavenumber $k_x$ by phase matching. Finally, the dispersion relation is $k^2=k_x^2 + k_z^2$ with $k_z=k\cos\theta$ and $k_x=k\sin\theta$. The top and bottom signs in~\eqref{eq:fields} correspond to an illumination propagating in the $+z$ or the $-z$ directions, respectively.

Due to the many structural symmetries that the scattering particle (dielectric cylinder) composing this metasurface exhibits, we know that $\te{\chi}_\text{em}=\te{\chi}_\text{me}=0$ and that $\te{\chi}_\text{ee}$ and $\te{\chi}_\text{mm}$ are diagonal matrices~\cite{angularAchouri2020,achouri2020fundamental}. Moreover, since we are only considering TM-polarized waves propagating in the $xz$-plane, the only susceptibility components that remain relevant to this problem are $\chi_\text{ee}^{xx}$, $\chi_\text{ee}^{zz}$, and $\chi_\text{mm}^{yy}$. This reduces~\eqref{eq:PM} to
\begin{subequations}\label{eq:PMsimple}
	\begin{align}
		P_x &= \epsilon_0\chi_\text{ee}^{xx} E_{x,\text{av}},\\
		P_z &= \epsilon_0\chi_\text{ee}^{zz} E_{z,\text{av}},\\
		M_y &= \chi_\text{mm}^{yy} H_{y,\text{av}}.
	\end{align}
\end{subequations}
Substituting~\eqref{eq:fields} and~\eqref{eq:PMsimple} into~\eqref{eq:GSTCsTD} and \eqref{eq:PM}, and solving the resulting system of equations for $\chi_\text{ee}^{xx}$ and $\chi_\text{mm}^{yy}$ in the case of an illumination propagating in the $+z$ direction at \textit{normal incidence}, i.e. $\theta=0^\circ$, as in Fig.~\ref{fig:MS}, yields
\begin{subequations}\label{eq:Xret}
	\begin{align}
        \chi_\text{ee}^{xx} &= \frac{2j}{k}\left(\frac{R_0-1+T_0}{R_0+1+T_0}\right),\\
		\chi_\text{mm}^{yy} &= \frac{2j}{k}\left(\frac{R_0+1-T_0}{R_0-1-T_0}\right),
	\end{align}
with subscript `0' indicating normal incidence. Repeating the process with (\ref{eq:GSTCsTD}b) using $\theta\neq 0^\circ$ yields
    \begin{align}\label{eq:Xretc}
        \chi_\text{ee}^{zz} &= \frac{2jk_z}{k_x^2}\left(\frac{R+1-T}{R-1-T}\right)-\frac{k^2}{k_x^2}\chi_\text{mm}^{yy} .
    \end{align}
\end{subequations}
In order to compute these susceptibilities, the metasurface reflection and transmission coefficients are now required. We obtain them from full-wave simulations for a metasurface whose physical parameters are provided in the caption of Fig.~\ref{fig:Limit}. These simulations are performed for an incidence angle ranging from $\theta=0^\circ$ to $\theta=85^\circ$ within the wavelength range 550~nm to 1500~nm (corresponding to a frequency range of 200~THz to 545~THz). The resulting transmitted power $(|T|^2)$ is plotted in Fig.~\ref{fig:Limit}a. 

The metasurface susceptibilities are now computed with~\eqref{eq:Xret} using the full-wave simulated reflection and transmission coefficients. The value of $\theta$ used to compute $\chi_\text{ee}^{zz}$ in~\eqref{eq:Xretc} is arbitrarily chosen to be $\theta=85^\circ$ to ensure that the metasurface normal polarization is sufficiently excited. The computed real part of these susceptibilities are plotted in Fig.~\ref{fig:Limit}b. Note that if the metasurface could really be described solely by these three dipolar susceptibilities, then the angle used to retrieve them would not matter. However, this is not the case here as modeling the full angular scattering response of this metasurface requires the introduction of higher-order susceptibility components, as shall be discussed in Sec.~\ref{sec:HOGSTC}. 

Now that we know the metasurface susceptibilities, we can use them to predict the angular scattering response of the metasurface and see if it indeed corresponds to the simulated result shown in Fig.~\ref{fig:Limit}a. To do so, we substitute~\eqref{eq:fields} and~\eqref{eq:PMsimple} into~\eqref{eq:GSTCsTD} and \eqref{eq:PM} and solve for the reflection and transmission coefficients; this yields
\begin{subequations}\label{eq:TRpred}
	\begin{align}
        R_\text{dip} &= \frac{2j\left(k^2\chi_\text{mm}^{yy}-k_z^2\chi_\text{ee}^{xx}+k_x^2\chi_\text{ee}^{zz} \right)}{\left(2j-k_z\chi_\text{ee}^{xx}\right)\left(k^2\chi_\text{mm}^{yy}+k_x^2\chi_\text{ee}^{zz}-2jk_z\right)},\\
		T_\text{dip} &= \frac{4k_z+k_z\chi_\text{ee}^{xx}\left(k_x^2\chi_\text{ee}^{zz}+k^2\chi_\text{mm}^{yy}\right)}{\left(2j-k_z\chi_\text{ee}^{xx}\right)\left(k^2\chi_\text{mm}^{yy}+k_x^2\chi_\text{ee}^{zz}-2jk_z\right)},
	\end{align}
\end{subequations}
where we use the subscript `dip', which stands for `dipolar', to avoid confusion with the simulated coefficients $R$ and $T$ used previously. Since a common practice in metasurface modeling has been to only consider the tangential susceptibility components and thus neglect the normal ones~\cite{angularAchouri2020}, we shall first evaluate the metasurface angular scattering response in the case where $\chi_\text{ee}^{zz}$ is purposefully set to zero and then look at the more general case where all three susceptibilities in~\eqref{eq:Xret} are nonzero. For this purpose, we now define, using the relations in~\eqref{eq:TRpred}, the purely tangential scattering parameters $R_\text{tang}=\left.R_\text{dip}\right|_{\chi_\text{ee}^{zz}=0}$ and $T_\text{tang}=\left.T_\text{dip}\right|_{\chi_\text{ee}^{zz}=0}$ with shorthand subscripts for `tangential'. Using the susceptibilities plotted in Fig.~\ref{fig:Limit}b and varying the incidence angle from $0^\circ$ to $85^\circ$ (which changes the value of $k_z$ in~\eqref{eq:TRpred}) results in the predicted transmission coefficient plotted in Fig.~\ref{fig:Limit}c. For comparison, we also plot the relative error between $|T|^2$ and $|T_\text{tang}|^2$ in Fig.~\ref{fig:Limit}d. As can be seen, this error remains acceptable for long wavelengths (where the period is optically small) and small incidence angles, which confirms that the modeling approach using only tangential susceptibilities is effective at least in the paraxial limit. However, the error becomes substantial as the wavelength becomes short, as well as at greater incidence angles.

In Fig.~\ref{fig:Limit}e-f, we plot the prediction and relative error using the complete dipolar susceptibilities from \eqref{eq:TRpred}. We see an improvement in the prediction at large incidence angles at long wavelengths ($>1200$ nm), thanks to the inclusion of the normal electric susceptibility. However, there is still a substantial error at shorter wavelengths, which clearly demonstrates the limitations of dipolar susceptibilities for modeling the general angular response of a metasurface.

\subsection{Multipolar Decomposition}
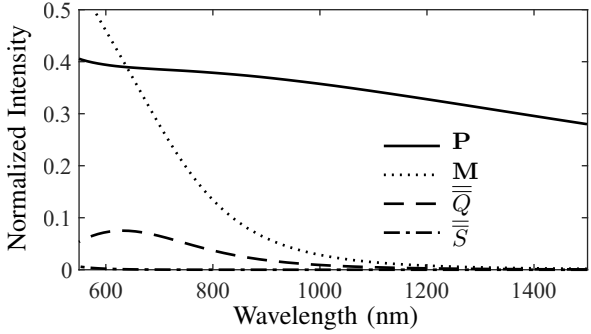
\begin{figure}[h!]
	\centering
	\begin{overpic}[width=\columnwidth,grid=false,trim={0cm 0cm 0cm 0cm},clip]{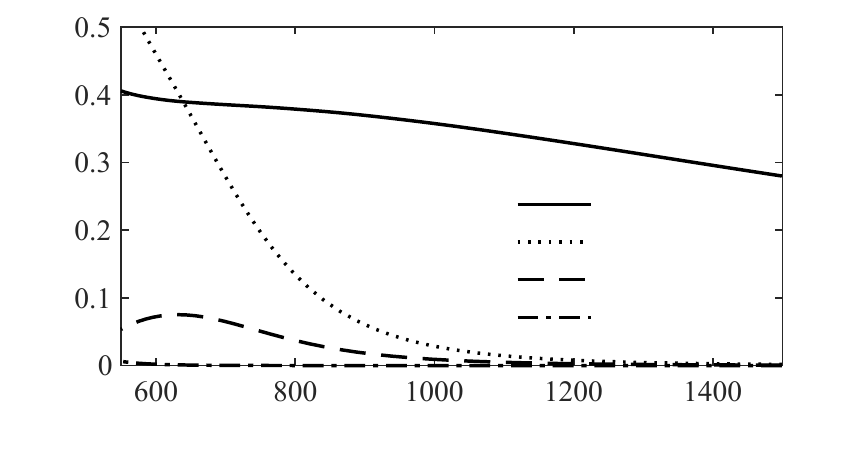}
	    \put(51,2){\htext{Wavelength (nm)}}
	    \put(5,29){\vtext{Normalized Intensity}}
	    \put(70,13.7){\small{}$\te{S}$}
	    \put(70,18.9){\small{}$\te{Q}$}
	    \put(70,23.8){\small{}$\ve{M}$}
	    \put(70,28.3){\small{}$\ve{P}$}
	\end{overpic}
	\caption{Normalized intensity of the far-field contributions from electric and magnetic dipolar ($\ve{P}$ and $\ve{M}$, respectively) and quadrupolar ($\te{Q}$ and $\te{S}$, respectively) moments induced in the metasurface considered in Fig.~\ref{fig:Limit}, with a normally-incident plane wave. The intensity of the field scattered by each component is plotted, normalized to the intensity of the incident field.}
	\label{fig:Decomposition}
\end{figure}

The issues with the dipolar modelling suggests the presence of higher-order multipolar polarization components. To verify that this is the case, a decomposition of the polarization currents\footnote{The current is obtained from full-wave simulations using \mbox{$\ve{J}=j\omega\epsilon_0(n^2-1)\ve{E}$}, where $n$ is the refractive index of the cylinders and $\ve{E}$ is the total field.} into Cartesian multipoles is performed, and subsequently the amplitude of the scattered fields from each multipole density is plotted in Fig~\ref{fig:Decomposition}. We follow the method from \cite{savinov_toroidal_2014}, having a unity-amplitude plane wave at normal incidence, and where toroidal contributions have been added to the corresponding primitive and traceless components. While the quadrupolar contributions to the scattered fields can be ignored at longer wavelengths, we see that there is a quadrupolar contribution that should not be ignored below $\sim 1000$ nm. 

In order to improve the angular scattering modeling of metasurfaces, we derive, in the next section, an extended version of the boundary conditions in~\eqref{eq:GSTCsTD} that includes higher-order multipolar components and spatially dispersive susceptibility tensors.

\section{Multipolar Modeling}
\label{sec:HOGSTC}

\subsection{Derivation of Higher-Order GSTCs}
\label{sec:Deriv}

We are now interested in deriving an expression of the GSTCs in~\eqref{eq:GSTCsTD} that include electric and magnetic dipolar and quadrupolar moments. We purposefully ignore higher-order multipolar moments for convenience but emphasize that the derivation that we next provide is easily extendable to any multipole moment.

Originally, the GSTCs were derived using a distribution-based approach where all quantities in Maxwell equations are expanded in terms of series of derivatives of the Dirac delta function~\cite{Idemen1973}. For instance, the expressions in~\eqref{eq:GSTCsTD} correspond to such a series expansion truncated at the 0th Dirac delta derivative order~\cite{Idemen1973, achouri2014general}. It is obviously possible to derive higher-order GSTCs by truncating these series at higher Dirac delta derivative orders, an example of which is provided in~\cite{achouri2015improvement} where the series are truncated at the 1st derivative order. 

An alternative derivation approach has been proposed in~\cite{albooyeh2016electromagnetic}, which consists in splitting Maxwell equations in tangential and normal components and then using conventional pillbox integration techniques to arrive at the GSTCs. However, this approach is also not well suited because it is not obvious how the electric and magnetic quadrupolar tensors should be split into tangential and normal parts.

While the distribution-based approach may be feasible, in this work the approach that we shall rather employ is based on the vector potential and was proposed in~\cite{silveirinhaBoundaryConditionsQuadrupolar2014} to derive boundary conditions that apply at the interface between media with quadrupolar moments. For our case, the method consists in computing the fields radiated by a metasurface using the vector potential, which may be related to a multipolar decomposition of the induced surface current on the metasurface. The GSTCs are then directly obtained by subtracting the fields on both sides of the metasurface. Since the derivation is based on the inhomogenous wave equation, it assumes identical media on either side of the metasurface, which is adequate for the freestanding metasurface considered in Fig.~\ref{fig:MS}. The benefit of this approach over the one based on distributions is its simplicity and accessibility.

Let us consider the vector potential $\ve{A}$, combined with the Lorenz gauge, from which the electric and magnetic fields are given by~\cite{jackson_classical_1999}
\begin{subequations}
    \label{eq:AtoEH}
	\begin{align}
		\ve{E} &= \frac{1}{j\omega\mu\epsilon}\left[\nabla\nabla + k^2\te{\text{I}} \right]\cdot\ve{A},\\
		\ve{H} &= \frac{1}{\mu}\nabla\times\ve{A}.
	\end{align}
\end{subequations}
In this approach, the metasurface is mathematically described by a fictitious sheet of electric current density $\ve{J}$, which forms, along with the vector potential $\ve{A}$, the inhomogeneous wave equation 
\begin{equation}
	\label{eq:AdiffEq}
	\nabla^2 \ve{A} + k^2 \ve{A} = -\mu \ve{J}.
\end{equation}
For a zero-thickness metasurface in the $xy$-plane at $z=0$, the current density may generally be expressed as
\begin{equation}
	\ve{J} = \delta(z) \ve{J}_\text{s}(x,y),
\end{equation}
where $\delta(z)$ is the Dirac delta function and $\ve{J}_\text{s}(x,y)$ is the metasurface spatially varying surface current density. Noting that $\ve{J}_\text{s}$ may be expressed in the spatial Fourier domain as
\begin{equation}
	\ve{J}_\text{s} = \int_{-\infty}^{+\infty}\int_{-\infty}^{+\infty} \ve{\tilde{J}_\text{s}}e^{-j(k_xx+k_yy)}dk_xdk_y,
\end{equation}
we may solve~\eqref{eq:AdiffEq} for $\ve{A}$, which yields\footnote{This solution is given for a single plane with wavevector $\ve{k}=(k_x,k_y,k_z)$ but may be easily generalized to any waveform by superposition. The validity of this solution may be verified by substituting~\eqref{eq:A} into~\eqref{eq:AdiffEq} and considering that $\partial_z^2 e^{-jk_z|z|} = -k_z^2e^{-jk_z |z|} - 2jk_z\delta(z)$.}
\begin{equation}\label{eq:A}
	\ve{A} = -\frac{j\mu}{2k_z}\ve{J}_\text{s}e^{-jk_z|z|}.
\end{equation}
The fields radiated by a metasurface with an arbitrary surface current distribution $\ve{J}_\text{s}$ are now obtained by substituting~\eqref{eq:A} into~\eqref{eq:AtoEH}, which leads to
\begin{subequations}\label{eq:MSfields}
	\begin{align}
		\ve{E} &= -\frac{1}{2k_z\omega\epsilon}\left[\nabla\nabla + k^2\te{\text{I}} \right]\cdot\ve{J}_\text{s}e^{-jk_z|z|},\\
		\ve{H} &= -\frac{j}{2k_z}\nabla\times\ve{J}_\text{s}e^{-jk_z|z|}.
	\end{align}
\end{subequations}
Since we are interested in expressing the GSTCs in terms of a multipolar expansion, we can replace $\ve{J}_\text{s}$ in~\eqref{eq:MSfields} by its multipolar expanded counterpart, which, truncated at the quadrupolar moments order, reads~\cite{papas2014theory,simovski2018composite,achouri2021extension}
\begin{equation}
	\label{eq:Jmulti}
	\ve{J}_\text{s} = j\omega\ve{P} + \nabla\times\ve{M} - \frac{j\omega}{2}\nabla\cdot\te{Q}  - \frac{1}{2}\nabla\times(\nabla\cdot\te{S}),
\end{equation}
where $\te{Q}$ and $\te{S}$ are the electric and magnetic quadrupolar moment densities, respectively. The extended GSTCs are now obtained by substituting~\eqref{eq:Jmulti} into~\eqref{eq:MSfields} and computing the differences of the fields between both sides of the metasurface with
\begin{subequations}\label{eq:delta}
	\begin{align}
		\Delta\ve{E} &= \ve{E}|_{|z|=z} - \ve{E}|_{|z|=-z},\\
		\Delta\ve{H} &= \ve{H}|_{|z|=z} - \ve{H}|_{|z|=-z}.
	\end{align}
\end{subequations}
After simplifying and rearranging the resulting expressions and setting $z=0$, we finally obtain the tangential components of the extended GSTCs as
\begin{subequations}\label{eq:GSTC_QS}
\begin{equation}
\begin{split}
	\hat{\ve{z}}\times\Delta\ve{E} &=  - j\omega\mu\ve{M}_\parallel + \frac{k^2}{2\epsilon}\ve{\hat{z}}\times \left(\te{Q}\cdot\ve{\hat{z}}\right) \\
	&\quad-\frac{1}{\epsilon}\hat{\ve{z}}\times\nabla\left[ P_z - \frac{1}{2}(\nabla_\parallel\ve{\hat{z}} + \ve{\hat{z}}\nabla_\parallel):\te{Q}\right] \\
	&\quad+ \frac{j\omega\mu}{2}\left(\te{S}- S_{zz}\te{I}\right)\cdot\nabla_\parallel,					
\end{split}		
\end{equation}\\
\begin{equation}
\begin{split}
	\hat{\ve{z}}\times\Delta\ve{H} &= j\omega\ve{P}_\parallel + \frac{k^2}{2}\ve{\hat{z}}\times \left(\te{S}\cdot\ve{\hat{z}}\right) \\
	&\quad- \hat{\ve{z}}\times \nabla\left[ M_z  -\frac{1}{2}(\nabla_\parallel\ve{\hat{z}}	+ \ve{\hat{z}}\nabla_\parallel):\te{S}\right] \\
	&\quad-\frac{j\omega}{2}\left(\te{Q}- Q_{zz}\te{I}\right)\cdot\nabla_\parallel.	
\end{split}	
\end{equation}
\end{subequations}
In these expressions, $Q_{zz} = \ve{\hat{z}}\cdot\te{Q}\cdot\ve{\hat{z}}$, $S_{zz} = \ve{\hat{z}}\cdot\te{S}\cdot\ve{\hat{z}}$, the dyadic $\nabla_\parallel\ve{\hat{z}} + \ve{\hat{z}}\nabla_\parallel$ is defined as
\begin{equation}\label{eq:Dyad}
	\nabla_\parallel\ve{\hat{z}} + \ve{\hat{z}}\nabla_\parallel =
	\begin{bmatrix}
		0 & 0 & \partial_x\\
		0 & 0 & \partial_y\\
		\partial_x & \partial_y & 0
	\end{bmatrix} ,
\end{equation}
and the double dot product between the arbitrary dyadics $\te{A}$ and $\te{B}$, as
\begin{equation}\label{key}
	\te{A}:\te{B} = A_{ij}B_{ij} = A_{11}B_{11} + A_{12}B_{12} + A_{13}B_{13} + \ldots
\end{equation}

\setcounter{equation}{24}
\begin{floatEq}
	\begin{equation}\label{eq:PMQS_simpl}
		\begin{bmatrix}
			P_x \\
			P_z\\
			M_y \\
			Q_{xx}\\
			Q_{xz} \\
			Q_{zz} \\
			S_{yz} \\
			S_{yx} 
		\end{bmatrix}
		\propto
		\begin{bmatrix}
			\chi_\text{ee}^{xx}  & \chi_\text{ee}^{xz}   &  \chi_\text{em}^{xy}   &  \chi_\text{ee}^{'xxx}   &  \chi_\text{ee}^{'xxz}  &  \chi_\text{ee}^{'xzz}  &  \chi_\text{em}^{'xyz}  &  \chi_\text{em}^{'xyx}  \\
			\chi_\text{ee}^{zx}  & \chi_\text{ee}^{zz}   &  \chi_\text{em}^{zy}   &  \chi_\text{ee}^{'zxx}   &  \chi_\text{ee}^{'zxz}  &  \chi_\text{ee}^{'zzz}  &  \chi_\text{em}^{'zyz}  &  \chi_\text{em}^{'zyx}  \\
			\chi_\text{me}^{yx}  & \chi_\text{me}^{yz}   &  \chi_\text{mm}^{yy}   &  \chi_\text{me}^{'yxx}   &  \chi_\text{me}^{'yxz}  &  \chi_\text{me}^{'yzz}  &  \chi_\text{mm}^{'yyz}  &  \chi_\text{mm}^{'yyx}  \\
			Q_\text{ee}^{xxx} & Q_\text{ee}^{xxz}  &  Q_\text{em}^{xxy}  &  Q_\text{ee}^{'xxxx}  &  Q_\text{ee}^{'xxxz}  &  Q_\text{ee}^{'xxzz}  &  Q_\text{em}^{'xxyz} &  Q_\text{em}^{'xxyx}  \\
			Q_\text{ee}^{xzx} & Q_\text{ee}^{xzz}  &  Q_\text{em}^{xzy}  &  Q_\text{ee}^{'xzxx}  &  Q_\text{ee}^{'xzxz}  &  Q_\text{ee}^{'xzzz}  &  Q_\text{em}^{'xzyz} &  Q_\text{em}^{'xzyx}  \\
			Q_\text{ee}^{zzx} & Q_\text{ee}^{zzz}  &  Q_\text{em}^{zzy}  &  Q_\text{ee}^{'zzxx}  &  Q_\text{ee}^{'zzxz}  &  Q_\text{ee}^{'zzzz}  &  Q_\text{em}^{'zzyz} &  Q_\text{em}^{'zzyx}  \\
			S_\text{me}^{yzx} & S_\text{me}^{yzz}  &  S_\text{mm}^{yzy}  &  S_\text{me}^{'yzxx}  &  S_\text{me}^{'yzxz}  &  S_\text{me}^{'yzzz}  &  S_\text{mm}^{'yzyz} &  S_\text{mm}^{'yzyx}  \\
			S_\text{me}^{yxx} & S_\text{me}^{yxz}  &  S_\text{mm}^{yxy}  &  S_\text{me}^{'yxxx}  &  S_\text{me}^{'yxxz}  &  S_\text{me}^{'yxzz}  &  S_\text{mm}^{'yxyz} &  S_\text{mm}^{'yxyx}  \\
		\end{bmatrix}
		\cdot
		\begin{bmatrix}
			E_{\text{av},x} \\
			E_{\text{av},z}\\
			H_{\text{av},y} \\
			\partial_xE_{\text{av},x}\\
			\partial_xE_{\text{av},z}/\partial_zE_{\text{av},x} \\
			\partial_zE_{\text{av},z} \\
			\partial_zH_{\text{av},y} \\
			\partial_xH_{\text{av},y}
		\end{bmatrix}
	\end{equation}
\end{floatEq}
\setcounter{equation}{18}

Now that we have derived GSTCs that include both dipolar and quadrupole moment densities, we have to properly express these moments in terms of the fields at the metasurface. In the case of relations~\eqref{eq:GSTCsTD}, which include only dipolar moments, the latter could be fully described in terms of the 4 bianisotropic susceptibility tensors in~\eqref{eq:PM}. However, by adding the electric and magnetic quadrupolar tensors in~\eqref{eq:GSTC_QS}, we now have the possibility to include many more susceptibility tensors by leveraging spatial dispersion, as shall be discussed in the next section.

\subsection{Spatial Dispersion}
\label{sec:SD}

Before providing the expressions of the moment densities that may be used in~\eqref{eq:GSTC_QS}, we shall first briefly review the concept of spatial dispersion. Fundamentally, spatial dispersion corresponds to a non-local response of a medium due to an exciting field. For instance, this implies that the induced current $\ve{J}$ in a medium due to the presence of an exciting electric field $\ve{E}$ may be expressed as the convolution~\cite{serdkov2001electromagnetics,simovski2018composite,achouri2021extension}
\begin{equation}
	\label{eq:WSDJ}
	\ve{J}(\ve{r}) = \int \te{K}(\ve{r}-\ve{r'})\cdot\ve{E}(\ve{r'})~dV',
\end{equation}
where $\te{K}$ represents the current response of the medium. By considering the three first terms of the Taylor expansion of $\ve{E}(\ve{r'})$ around $\ve{r}$, we may transform~\eqref{eq:WSDJ} into~\cite{serdkov2001electromagnetics,simovski2018composite,achouri2021extension}
\begin{equation}
	\label{eq:Jexp}
	J_i = b_{ij}E_j + b_{ijk}\nabla_kE_j + b_{ijkl}\nabla_l\nabla_kE_j,
\end{equation}
where the tensor $b_{ij}$ represents a local response of the medium to the exciting electric field, while the tensors $b_{ijk}$ and $b_{ijkl}$ represent its first- and second-order nonlocal responses, respectively. 

The tensors in~\eqref{eq:Jexp} are conventionally split into symmetric and antisymmetric parts so as to associate them to electric and magnetic excitations, respectively. For instance, by splitting $b_{ijk}$ into its symmetric $(b_{ijk}^\text{sym})$ and antisymmetric ($b_{ijk}^\text{asym}$) parts, the second term on the right-hand side of~\eqref{eq:Jexp} may be expressed as
\begin{equation}
	\label{eq:bsplit}
	\begin{split}
		b_{ijk}\nabla_kE_j &= \left(b_{ijk}^\text{sym}+b_{ijk}^\text{asym}\right)\nabla_kE_j,\\
		&=\left(b_{ijk}^\text{sym} + \frac{j}{\omega\mu_0}\varepsilon_{ljk}g_{il}\right)\nabla_kE_j,\\
		&=b_{ijk}^\text{sym}\nabla_kE_j + g_{ij}H_j,
	\end{split}
\end{equation}
where we have used the fact that the antisymmetric third-rank tensor $b_{ijk}^\text{asym}$ may be equivalently expressed as the second-rank tensor $g_{il}$ using the Levi-Civita symbol $\varepsilon_{ijk}$~\cite{arfken1999mathematical}. In the last part of~\eqref{eq:bsplit}, we have also used the fact that the Maxwell equation $\nabla\times\ve{E} = -j\omega\mu_0 \ve{H}$ may be written as $\varepsilon_{ijk}\nabla_kE_j=-j\omega\mu_0 H_i$. The decomposition in~\eqref{eq:bsplit} explains, for instance, the dependence of $\ve{P}$ on $\ve{H}$ via the susceptibility tensor $\te{\chi}_\text{em}$ in~\eqref{eq:PM}.

Noting that the steps from~\eqref{eq:WSDJ} to~\eqref{eq:bsplit} may be repeated for $\ve{P}$, $\ve{M}$, $\te{Q}$ and $\te{S}$, we may now express the extension of~\eqref{eq:PM} as 
\begin{equation}
	\label{eq:PMQS}
	\begin{bmatrix}
		P_i\\
		M_i\\
		Q_{il}\\
		S_{il}
	\end{bmatrix}=
	\te{\chi}\cdot
	\begin{bmatrix}
		E_{\text{av},j}\\
		H_{\text{av},j}\\
		\nabla_k E_{\text{av},j}\\
		\nabla_k H_{\text{av},j}
	\end{bmatrix},
\end{equation}
where the hypersusceptibility tensor $\te{\chi}$ is given by
\begin{equation}
	\label{eq:chi}
	\te{\chi}=
	\begin{bmatrix}
		\epsilon_0\chi_{\text{ee}}^{ij} & \frac{1}{c_0}\chi_{\text{em}}^{ij} & \frac{\epsilon_0}{2k_0}\chi_{\text{ee}}^{'ijk} & \frac{1}{2c_0k_0}\chi_{\text{em}}^{'ijk}\\
		\frac{1}{\eta_0}\chi_{\text{me}}^{ij} & \chi_{\text{mm}}^{ij} & \frac{1}{2\eta_0k_0}\chi_{\text{me}}^{'ijk} & \frac{1}{2k_0}\chi_{\text{mm}}^{'ijk}\\
		\frac{\epsilon_0}{k_0}Q_{\text{ee}}^{ilj} & \frac{1}{c_0k_0}Q_{\text{em}}^{ilj} & \frac{\epsilon_0}{2k_0^2}Q_{\text{ee}}^{'iljk} & \frac{1}{2c_0k_0^2}Q_{\text{em}}^{'iljk}\\
		\frac{1}{\eta_0k_0}S_{\text{me}}^{ilj} & \frac{1}{k_0}S_{\text{mm}}^{ilj} & \frac{1}{2\eta_0k_0^2 }S_{\text{me}}^{'iljk} & \frac{1}{2k_0^2}S_{\text{mm}}^{'iljk}
	\end{bmatrix}.
\end{equation}
In these expressions, we retrieve the 4 bianisotropic susceptibility tensors that were already present in~\eqref{eq:PM} and gain several additional tensors that appear due to spatial dispersion and the presence of the quadrupolar moment densities. The constants present in~\eqref{eq:chi} are used to ensure that all surface tensors have units expressed in (m). We emphasize that the tensorial components in~\eqref{eq:chi} are not all independent from each. Indeed, one \textit{must} consider the symmetry relations~\eqref{eq:sym}, which \textit{always} apply, and the reciprocity conditions~\eqref{eq:recip1} and~\eqref{eq:recip2}, which apply if the metasurface is reciprocal. 

A striking feature that results from the reciprocity conditions is that, for instance, the dependence of $\ve{P}$ on the gradient of the electric field (via $\chi'_{\text{ee},ijk}$) is reciprocally connected to the dependence of $\te{Q}$ on the electric field (via $Q_{\text{ee},ijk}$). This means that in modeling a reciprocal metasurface, it would not make sense to include $\chi'_{\text{ee},ijk}$ without also including $Q_{\text{ee},ijk}$ and thus using boundary conditions that include $\te{Q}$. This is why in~\eqref{eq:PM} we can safely consider only the 4 bianisotropic susceptibility tensors without including higher-order terms since these equations are limited to the dipolar moments. By the same token, the symmetric part of $b_{ijkl}$ in~\eqref{eq:Jexp}, which is related to $\nabla_l\nabla_kE_j$, would be reciprocally connected to components of the electric octupole moment. Since the octupole moments are not taken into account in~\eqref{eq:GSTC_QS}, we do not include the components related to $\nabla_l\nabla_kE_j$ in~\eqref{eq:PMQS}. However, the antisymmetric part of $b_{ijkl}$, which may be expressed in terms of $\nabla_k H_j$, is reciprocally related to the magnetic quadrupolar moment $\te{S}$, hence its presence in~\eqref{eq:PMQS}.

\begin{figure*}[t!]
	\centering
	\includegraphics[width=0.75\linewidth]{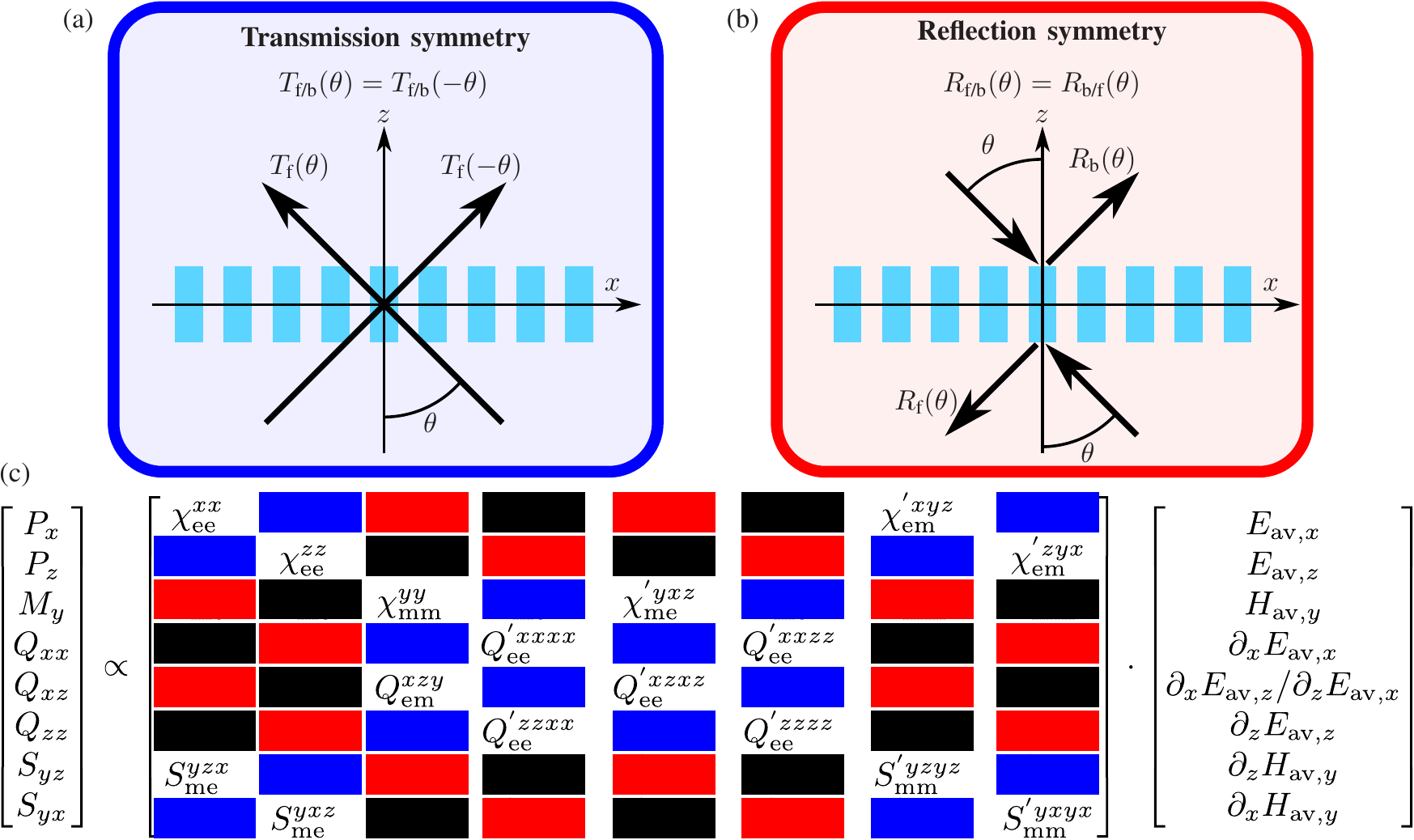}
	\caption{Scattering symmetries and their impacts on the metasurface modeling. (a)~Transmission symmetry. (b)~Reflection symmetry. (c)~Hypersusceptibilities playing a role in the scattering of TM-polarized waves propagating in the $xz$-plane. The blue and red rectangles indicate hypersusceptibilities that lead to asymmetric transmission and reflection, respectively, whereas the black rectangles correspond to hypersusceptibilities that self-cancel due to reciprocity.}
	\label{fig:matrix}
\end{figure*}

\subsection{Illustrative Example}
\label{sec:example}

We shall now come back to the modeling problem discussed in Sec.~\ref{sec:lim} and apply the newly derived extended GSTCs~\eqref{eq:GSTC_QS} and the associated spatially-dispersive moments~\eqref{eq:PMQS}.

In this problem, we are only interested in modeling the interactions of TM-polarized waves propagating in the $xz$-plane, so we can reduce the GSTCs in~\eqref{eq:GSTC_QS} to
\begin{subequations}\label{eq:Simplified_GSTC}
	\begin{align}
		\begin{split}\label{eq:Simplified_GSTC1}
				\Delta E_x &= \frac{j\omega\mu}{2} \left(\partial_x S_{yx} - 2M_y\right)+\frac{1}{2\epsilon} \partial_x^2(Q_{zx}+Q_{xz})\\
				&\qquad+\frac{\omega k^2}{2\epsilon} Q_{xz}-\frac{1}{\epsilon} \partial_xP_z,
		\end{split}\\
	\Delta H_y &= -\frac{1}{2}\left[2j\omega P_x  - k^2S_{yz} -j\partial_x\omega (Q_{xx}-Q_{zz})\right].
	\end{align}
\end{subequations}
Similarly, we reduce~\eqref{eq:PMQS} so that the multipolar moments in~\eqref{eq:Simplified_GSTC} are only expressed in terms of the field components $E_x$, $E_z$ and $H_y$ as well as their derivatives along $x$ and $z$. The resulting expression is provided in~\eqref{eq:PMQS_simpl}, where we have not included all the prefactors in~\eqref{eq:chi} for convenience. Note that the symmetry relations~\eqref{eq:sym} are already taken into account in~\eqref{eq:PMQS_simpl}.

To model the metasurface in Fig.~\ref{fig:MS}, we now follow the same procedure as in Sec.~\ref{sec:lim}, i.e., we aim at finding an expression equivalent to~\eqref{eq:Xret} using~\eqref{eq:PMQS_simpl} instead of~\eqref{eq:PMsimple} and~\eqref{eq:Simplified_GSTC} instead of~\eqref{eq:GSTCsTD}. The problem is obviously more complicated now since there are 64 components in~\eqref{eq:PMQS_simpl}, whereas there was only 2 (or 3 if $\chi_\text{ee}^{zz}$ is considered) components in~\eqref{eq:PMsimple}. However, since we are interested in developing a model that is physically sound, and thus consistent with the electromagnetic response of the metasurface scattering particles, we will see that most of these components either do not contribute or are in fact dependent on each other.

To simplify~\eqref{eq:PMQS_simpl}, the first fundamental concept that must be considered is reciprocity. We know that the metasurface in Fig.~\ref{fig:MS} is reciprocal since we do not bias it with a time-odd external quantity~\cite{kong1986electromagnetic,Jackson1998,calozElectromagneticNonreciprocity2018}. Therefore, the conditions~\eqref{eq:recip1} and~\eqref{eq:recip2} must be satisfied implying that the lower triangular part of~\eqref{eq:PMQS_simpl} depends on its upper triangular part. This reduces the number of independent unknowns from 64 to 36. Moreover, reciprocity also implies that some components in~\eqref{eq:PMQS_simpl} cancel each other when substituted in~\eqref{eq:Simplified_GSTC}. Indeed, this is for instance the case of $\chi_\text{em}^{zy}$ and $\chi_\text{me}^{yz}$, which are connected to each other by reciprocity and that end up canceling each other when substituted in~\eqref{eq:Simplified_GSTC1} via $P_z$ and $M_y$, respectively. All the terms that cancel each other like this are highlighted by a black rectangle in Fig.~\ref{fig:matrix}c. This further reduces the number of independent unknowns in~\eqref{eq:PMQS_simpl} to 28.

The second concept that must be considered is the angular response of the metasurface scattering particle. The one that is used in our case is a simple cylinder that exhibits 2 particularly important symmetries. Among others, it is mirror symmetric through the $yz$ and the $xy$ planes. The mirror symmetry through the $yz$-plane implies that its angular transmission coefficient in the $xz$-plane is symmetric with respect to $\theta$~\cite{angularAchouri2020}, i.e., $T_\text{f/b}(\theta) = T_\text{f/b}(-\theta)$ where `f' and `b' stand for forward ($+z$) and backward ($-z$) illumination directions, respectively. Similarly, the mirror symmetry through the $xy$-plane implies that its angular reflection coefficient is the same for illuminations impinging on either sides of the metasurface~\cite{angularAchouri2020}, i.e., $R_\text{f/b}(\theta) = R_\text{b/f}(\theta)$. These two cases of symmetric angular transmission and reflection are depicted in Figs.~\ref{fig:matrix}a and~\ref{fig:matrix}b, respectively. Now, it turns out that some of the components in~\eqref{eq:PMQS_simpl} lead to \textit{asymmetric} angular scattering. Therefore, these components must not be taken into account in our case because they would lead to an unphysical model. To find out which components lead to asymmetric angular scattering, we substitute~\eqref{eq:fields} into~\eqref{eq:Simplified_GSTC} and investigate the scattering response of each of the remaining 28 independent components in~\eqref{eq:PMQS_simpl} individually. Those who do not satisfy the transmission symmetry condition in Fig.~\ref{fig:matrix}a are highlighted by a blue rectangle in Fig.~\ref{fig:matrix}c, whereas those who do not satisfy the reflection symmetry condition in Fig.~\ref{fig:matrix}b are highlighted by a red rectangle. We are now left with only 12 independent and relevant components in~\eqref{eq:PMQS_simpl}.

We now further simplify~\eqref{eq:Simplified_GSTC} by using~\eqref{eq:fields} along with the 12 independent components that are left in~\eqref{eq:PMQS_simpl}, which leads to
\stepcounter{equation}
\begin{subequations}\label{eq:RTPMQS}
	\begin{align}
		\begin{split}
			\frac{2}{jk}\frac{(1+R-T)}{(1-R+T)} &= \sec(\theta)A  +\sin(\theta)\tan(\theta)B  \\
			&\quad+\frac{1}{4}\cos^2 (2\theta)\sec(\theta){Q}_\text{ee}^{'xzxz},
		\end{split}\label{eq:RTPMQS1}\\
		\frac{2}{jk}\frac{(1-R-T)}{(1+R+T)} &= \cos(\theta)C +\frac{1}{4}\cos(\theta)\sin^2(\theta)D,\label{eq:RTPMQS2}
	\end{align}
\end{subequations}
where $k_x$ and $k_z$ have been replaced by $k_x=k\sin\theta$ and  $k_z=k\cos\theta$ for convenience. In simplifying these expressions, we have grouped together several of the remaining components in~\eqref{eq:PMQS_simpl} because they exhibit the same angular scattering response. With the exception of ${Q}_\text{ee}^{'xzxz}$, which presents a unique angular scattering response, all the other terms have been replaced by arbitrarily named variables that are defined as
\begin{subequations}\label{eq:tilde}
	\begin{align}	
		A &= \chi_\text{mm}^{yy}  -j\chi_\text{me}^{'yxz},\label{eq:tilde1}\\
		B &= \chi_\text{ee}^{zz}+j\chi_\text{em}^{'zyx}+\frac{1}{4}S_\text{mm}^{'yxyx}+2j\chi_\text{me}^{'yxz},\\
		C &= \chi_\text{ee}^{xx} -j\chi_\text{em}^{'xyz}+\frac{1}{4}S_\text{mm}^{'yzyz},\\
		D &= Q_\text{ee}^{'xxzz} -\frac{1}{2}Q_\text{ee}^{'xxxx}-\frac{1}{2}Q_\text{ee}^{'zzzz}.
	\end{align}
\end{subequations}
This means that even though there are 12 independent unknowns that could, in principle, be solved for in~\eqref{eq:PMQS_simpl}, we cannot compute them all since some of them have identical effects on the metasurface scattering response and would thus lead to an ill-conditioned system of equations\footnote{While the method that we propose here is unable to retrieve all the independent components in~\eqref{eq:PMQS_simpl}, it may be possible to compute them using the alternative technique proposed in~\cite{Koendrink2014Retrieval}, which consists in selectively canceling the fields or their derivatives at the metasurface to excite only the desired components.}. We shall instead restrict our attention to the 5 unknowns that are left in~\eqref{eq:RTPMQS}. To obtain them, we note that the system~\eqref{eq:RTPMQS} is in fact made of two independent equations that can thus be solved individually. Accordingly, Eq.~\eqref{eq:RTPMQS1}, which contains 3 unknowns, is solved by using the simulated metasurface reflection and transmission coefficients at 3 different angles of incidence, whereas Eq.~\eqref{eq:RTPMQS2}, which contains 2 unknowns, is solved by using only 2 different angles of incidence.
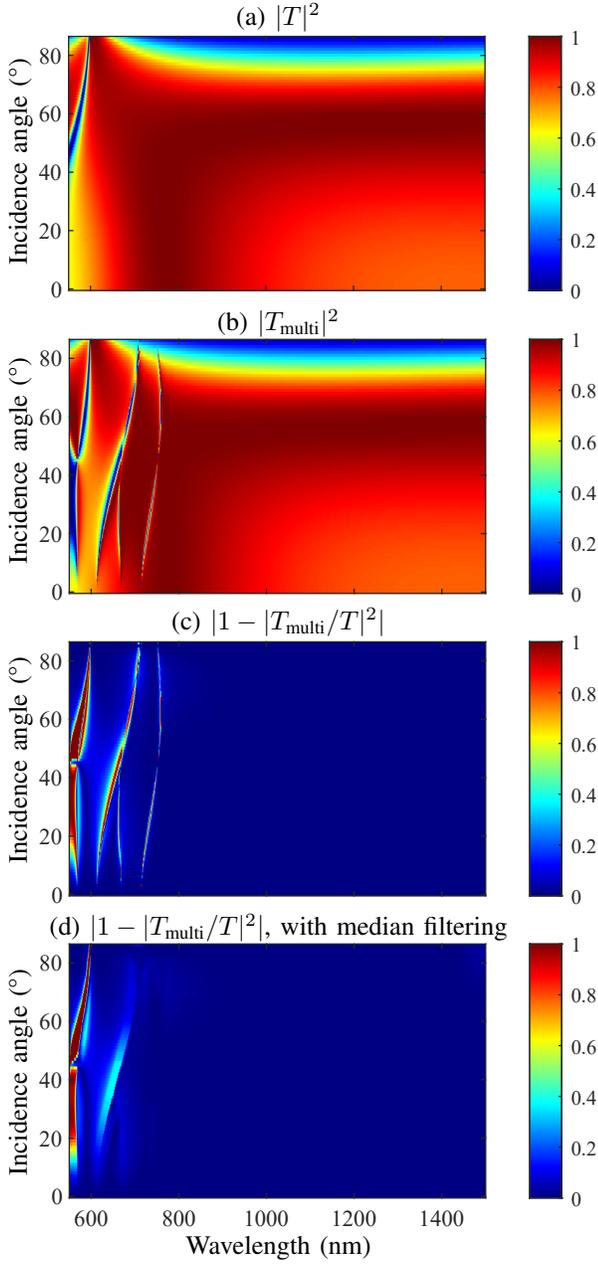
\begin{figure}[h!]
	\centering
    \begin{overpic}[width=\columnwidth,grid=false,trim={0cm 0cm 0cm 0cm},clip]{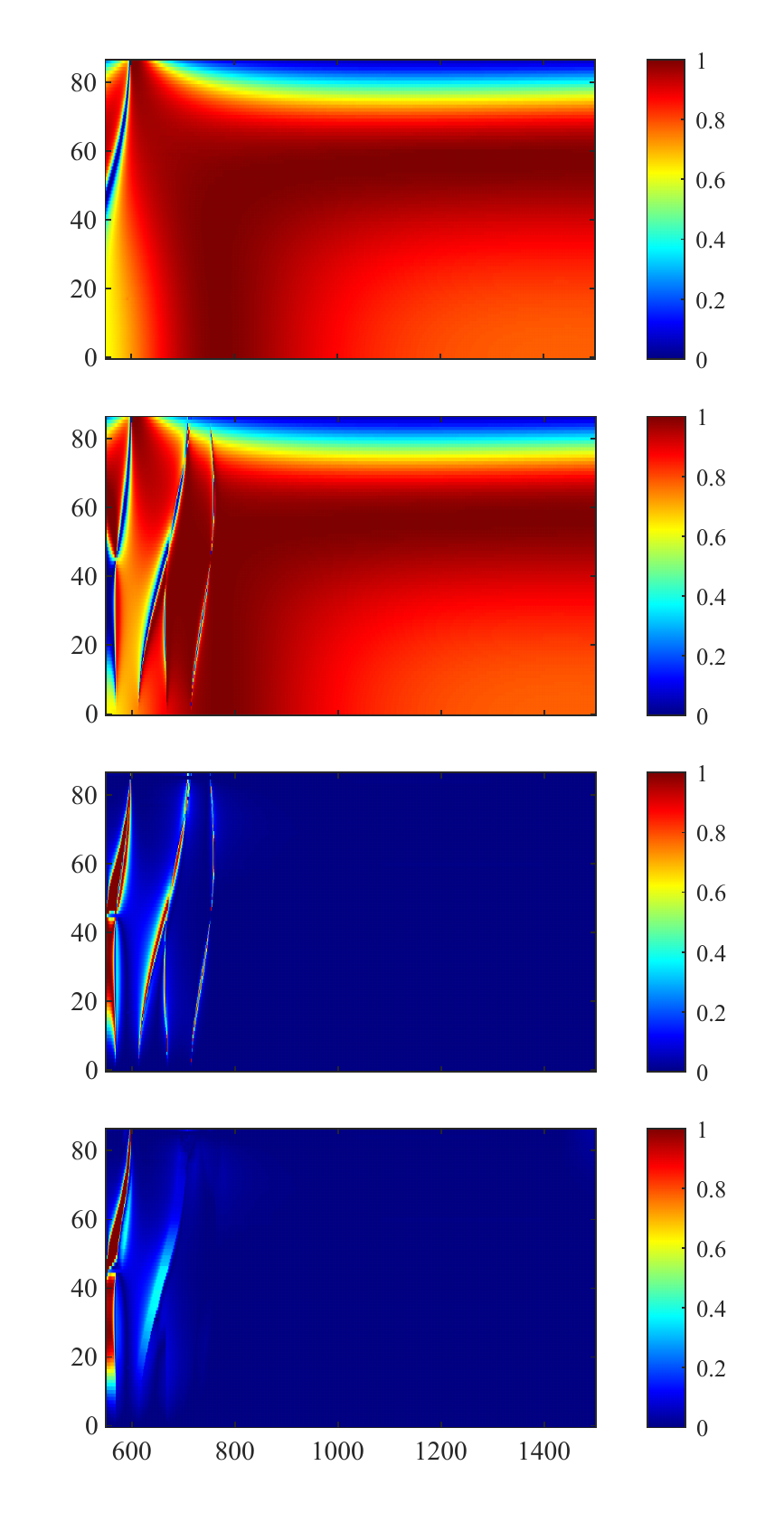}
    	    \put(23,2.5){\htext{Wavelength (nm)}}
    	    \put(3.2,16){\vtext{Incidence angle (°)}}
    	    \put(3.2,40){\vtext{Incidence angle (°)}}
    	    \put(3.2,63){\vtext{Incidence angle (°)}}
    	    \put(3.2,86){\vtext{Incidence angle (°)}}
    	    \put(23,97){\htext{(a) $|T|^2$}}
    	    \put(23,73.5){\htext{(b) $|T_\text{multi}|^2$}}
    	    \put(23,50.5){\htext{(c) $|1-|T_\text{multi}/T|^2|$}}
    	    \put(23,27){\htext{(d) $|1-|T_\text{multi}/T|^2|$, with median filtering}}
    \end{overpic}
	\caption{Multipolar modeling of the metasurface in Fig.~\ref{fig:MS} with $P=225$~nm, $D=200$~nm, $H=400$~nm and $n=2.55$. (a) Simulated transmission coefficient. (b) Predicted transmission coefficient. (c) Relative transmission error. (c) Relative transmission error after processing $|T_\text{multi}|^2$ using a median filter having a width of 30 nm.}
	\label{fig:PMQS_pred}
\end{figure}

To provide the system of equations with enough information about the angular response of the metasurface, we select the 3 angles for~\eqref{eq:RTPMQS1} to be $\theta=[0^\circ, 45^\circ, 85^\circ]$, and the 2 angles for~\eqref{eq:RTPMQS2} to be $\theta=[0^\circ, 85^\circ]$. Once the 5 unknowns in~\eqref{eq:RTPMQS} are obtained, we then reverse these equations to predict the angular reflection ($R_\text{multi}$) and transmission ($T_\text{multi}$) coefficients for all other angles. The resulting predicted transmission coefficient is plotted in Fig.~\ref{fig:PMQS_pred}b, and the relative error between the simulated data shown in Fig.~\ref{fig:PMQS_pred}a (repeated from Fig.~\ref{fig:Limit}a) and the predicted one is plotted in Fig.~\ref{fig:PMQS_pred}c.

Beside the appearance of some undesired sharp features, the overall predicted transmission coefficient is in much better agreement with the simulated data in Fig.~\ref{fig:PMQS_pred}b than it is in Fig.~\ref{fig:Limit}e, especially for large angles of incidence. In fact, the total error, defined as the sum of the difference between simulated and predicted data for each combination of incidence angle and wavelength point from 600 nm to 1500 nm, is 3.5 times smaller when using the multipolar modeling approach than the purely dipolar one (Fig.~\ref{fig:Limit}e). 
While the agreement between the simulated data and multipolar modelling is good up above about 760 nm, the sharp vertical features can be alleviated through the application of a median filter. By filtering $|T_\text{multi}|^2$ with a window of 30 nm, the resulting error is plotted in Fig.~\ref{fig:PMQS_pred}c. The total error in this case is 6.1 times smaller than the dipolar approach, over the span from 600 nm to 1500 nm.

\section{Conclusions}
\label{sec:concl}

We have shown that the conventional GSTCs, which only include dipolar moments, are rather limited when it comes to modeling the angular scattering response of a metasurface, especially in the case of high frequencies (small wavelength-to-period ratios) and/or large incidence angles. In order to improve this model, we have extended it by adding quadrupolar moments and a plethora of associated susceptibility tensors that result from the presence of spatially dispersive effects. Our derivation applies for a metasurface in a uniform environment. The extension of the proposed boundary conditions to the more general case of different media on either side, e.g. the presence of a substrate, remains future work.

We have demonstrated that the derived multipolar GSTCs provide an improvement by at least a factor of 3.5 in the modeling accuracy of the angular response of metasurfaces. Note that, in the proposed example, this accuracy improvement was achieved by only considering 5 effective (hyper)susceptibilities, which is only 3 more than the number of susceptibilities used in the dipolar model. Taking into account more terms or higher-order multipolar components should most like help providing an even better accuracy.

In addition, the presence of the large number of hypersusceptibility components in the multipolar GSTCs may not be only attractive for improving its modeling accuracy but also for providing many new degrees of freedom for controlling electromagnetic waves. The ability to engineer not only the dipolar but also the quadrupolar responses of a metasurface is thus expected to be instrumental in the design of multi-functional angular processing metasurfaces.

\appendices

\section{Symmetry Relations}
\label{sec:symm}

Due to the way the quadrupolar tensor $\te{Q}$ is defined and due to the symmetric-antisymmetric splitting in~\eqref{eq:bsplit}, the following symmetry relations apply~\cite{LightPropagationCubic1990,Jackson1998,simovski2018composite,achouri2021extension}
\begin{equation}
	\label{eq:sym}
	\begin{split}
		Q_{\text{ee},ijk} &= Q_{\text{ee},jik},\quad Q_{\text{em},ijk} = Q_{\text{em},jik},\\
		\chi'_{\text{ee},ijk} &= \chi'_{\text{ee},ikj},\quad \chi'_{\text{me},ijk} = \chi'_{\text{me},ikj},\\
		Q'_{\text{em},ijkl} &= Q'_{\text{em},jikl},\quad S'_{\text{me},ijkl} = S'_{\text{me},ijlk},\\
		Q'_{\text{ee},ijkl} &= Q'_{\text{ee},jikl} = Q'_{\text{ee},ijlk} = Q'_{\text{ee},jilk}.
	\end{split}
\end{equation}	

\section{Reciprocity Relations}
\label{sec:recip}

The reciprocity relations for a bianisotropic medium are~\cite{kong1986electromagnetic,rothwell2018electromagnetics,calozElectromagneticNonreciprocity2018}
\begin{equation}
	\label{eq:recip1}
	\chi_{\text{ee},ij} = \chi_{\text{ee},ji}, \quad \chi_{\text{mm},ij} = \chi_{\text{mm},ji}, \quad \chi_{\text{em},ij} = -\chi_{\text{me},ji},
\end{equation}
while those for higher-order components are given by~\cite{LightPropagationCubic1990,simovski2018composite,achouri2021extension}
\begin{equation}
	\label{eq:recip2}
	\begin{split}
		\chi'_{\text{ee},kji} &= Q_{\text{ee},ijk},\quad  \chi'_{\text{mm},kij} = S_{\text{mm},ijk},\\
		\chi'_{\text{em},kij} &=- S_{\text{me},ijk},\quad \chi'_{\text{me},kji} = -Q_{\text{em},ijk},\\
		Q'_{\text{ee},klij} &= Q'_{\text{ee},ijkl},\quad Q'_{\text{em},klij} = - S'_{\text{me},ijkl},\\
		&~\quad S'_{\text{mm},klij}=S'_{\text{mm},ijkl}.	
	\end{split}
\end{equation}

\bibliographystyle{myIEEEtran}
\bibliography{NewLib}

\begin{thebibliography}{10}
\providecommand{\url}[1]{#1}
\csname url@samestyle\endcsname
\providecommand{\newblock}{\relax}
\providecommand{\bibinfo}[2]{#2}
\providecommand{\BIBentrySTDinterwordspacing}{\spaceskip=0pt\relax}
\providecommand{\BIBentryALTinterwordstretchfactor}{4}
\providecommand{\BIBentryALTinterwordspacing}{\spaceskip=\fontdimen2\font plus
\BIBentryALTinterwordstretchfactor\fontdimen3\font minus
  \fontdimen4\font\relax}
\providecommand{\BIBforeignlanguage}[2]{{%
\expandafter\ifx\csname l@#1\endcsname\relax
\typeout{** WARNING: IEEEtran.bst: No hyphenation pattern has been}%
\typeout{** loaded for the language `#1'. Using the pattern for}%
\typeout{** the default language instead.}%
\else
\language=\csname l@#1\endcsname
\fi
#2}}
\providecommand{\BIBdecl}{\relax}
\BIBdecl

\bibitem{landy2008perfect}
N.~I. Landy, S.~Sajuyigbe, J.~J. Mock, D.~R. Smith, and W.~J. Padilla,
  ``Perfect metamaterial absorber,'' \emph{Physical review letters}, vol. 100,
  no.~20, p. 207402, 2008.

\bibitem{schurig2006metamaterial}
D.~Schurig, J.~J. Mock, B.~Justice, S.~A. Cummer, J.~B. Pendry, A.~F. Starr,
  and D.~R. Smith, ``Metamaterial electromagnetic cloak at microwave
  frequencies,'' \emph{Science}, vol. 314, no. 5801, pp. 977--980, 2006.

\bibitem{kildishev2013planar}
A.~V. Kildishev, A.~Boltasseva, and V.~M. Shalaev, ``Planar photonics with
  metasurfaces,'' \emph{Science}, vol. 339, no. 6125, 2013.

\bibitem{yu2014flat}
N.~Yu and F.~Capasso, ``Flat optics with designer metasurfaces,'' \emph{Nature
  Mater.}, vol.~13, no.~2, pp. 139--150, 2014.

\bibitem{chenHuygensMetasurfacesMicrowaves2018}
M.~Chen, M.~Kim, A.~M. Wong, and G.~V. Eleftheriades, ``Huygens' metasurfaces
  from microwaves to optics: A review,'' \emph{Nanophotonics}, vol.~7, no.~6,
  pp. 1207--1231, Jun. 2018.

\bibitem{achouri2020electromagnetic}
K.~Achouri and C.~Caloz, \emph{Electromagnetic Metasurfaces: Theory and
  Applications}.\hskip 1em plus 0.5em minus 0.4em\relax Wiley-IEEE Press, 2021.

\bibitem{6477089}
T.~Niemi, A.~Karilainen, and S.~Tretyakov, ``Synthesis of polarization
  transformers,'' \emph{IEEE Trans. Antennas Propag.}, vol.~61, no.~6, pp.
  3102--3111, June 2013.

\bibitem{achouri2014general}
K.~Achouri, M.~A. Salem, and C.~Caloz, ``General metasurface synthesis based on
  susceptibility tensors,'' \emph{IEEE Trans. Antennas Propag.}, vol.~63,
  no.~7, pp. 2977--2991, Jul. 2015.

\bibitem{epstein2016arbitrary}
A.~Epstein and G.~Eleftheriades, ``Passive lossless {H}uygens metasurfaces for
  conversion of arbitrary source field to directive radiation,'' \emph{IEEE
  Trans. Antennas Propag.}, vol.~62, no.~11, pp. 5680--5695, Nov 2014.

\bibitem{diazrubioGeneralizedReflectionLaw2017}
A.~{D{\'i}az-Rubio}, V.~S. Asadchy, A.~Elsakka, and S.~A. Tretyakov, ``From the
  generalized reflection law to the realization of perfect anomalous
  reflectors,'' \emph{SCIENCE ADVANCES}, p.~11, 2017.

\bibitem{Lavigne2018}
G.~Lavigne, K.~Achouri, V.~S. Asadchy, S.~A. Tretyakov, and C.~Caloz,
  ``Susceptibility derivation and experimental demonstration of refracting
  metasurfaces without spurious diffraction,'' \emph{IEEE Trans. Antennas
  Propag.}, vol.~66, no.~3, pp. 1321--1330, March 2018.

\bibitem{porsAnalogComputingUsing2015}
A.~Pors, M.~G. Nielsen, and S.~I. Bozhevolnyi, ``Analog {{Computing Using
  Reflective Plasmonic Metasurfaces}},'' \emph{Nano Letters}, vol.~15, no.~1,
  pp. 791--797, Jan. 2015.

\bibitem{chizariAnalogOpticalComputing2016}
A.~Chizari, S.~Abdollahramezani, M.~V. Jamali, and J.~A. Salehi, ``Analog
  optical computing based on a dielectric meta-reflect array,'' \emph{Optics
  Letters}, vol.~41, no.~15, p. 3451, Aug. 2016.

\bibitem{abdollahramezaniDielectricMetasurfacesSolve2017}
S.~Abdollahramezani, A.~Chizari, A.~E. Dorche, M.~V. Jamali, and J.~A. Salehi,
  ``Dielectric metasurfaces solve differential and integro-differential
  equations,'' \emph{Optics Letters}, vol.~42, no.~7, p. 1197, Apr. 2017.

\bibitem{babashahIntegrationAnalogOptical2017}
H.~Babashah, Z.~Kavehvash, S.~Koohi, and A.~Khavasi, ``Integration in analog
  optical computing using metasurfaces revisited: Toward ideal optical
  integration,'' \emph{Journal of the Optical Society of America B}, vol.~34,
  no.~6, p. 1270, Jun. 2017.

\bibitem{zhuPlasmonicComputingSpatial2017}
T.~Zhu, Y.~Zhou, Y.~Lou, H.~Ye, M.~Qiu, Z.~Ruan, and S.~Fan, ``Plasmonic
  computing of spatial differentiation,'' \emph{Nature Communications}, vol.~8,
  no.~1, p. 15391, Aug. 2017.

\bibitem{momeni2019generalized}
A.~Momeni, H.~Rajabalipanah, A.~Abdolali, and K.~Achouri, ``Generalized optical
  signal processing based on multioperator metasurfaces synthesized by
  susceptibility tensors,'' \emph{Physical Review Applied}, vol.~11, no.~6, p.
  064042, 2019.

\bibitem{zhangControllingAngularDispersions2020}
X.~Zhang, Q.~Li, F.~Liu, M.~Qiu, S.~Sun, Q.~He, and L.~Zhou, ``Controlling
  angular dispersions in optical metasurfaces,'' \emph{Light: Science \&
  Applications}, vol.~9, no.~1, p.~76, Dec. 2020.

\bibitem{angularAchouri2020}
K.~{Achouri} and O.~J.~F. {Martin}, ``Angular scattering properties of
  metasurfaces,'' \emph{IEEE Transactions on Antennas and Propagation},
  vol.~68, no.~1, pp. 432--442, 2020.

\bibitem{raabMultipoleTheoryElectromagnetism2005}
R.~E. Raab and O.~L. De~Lange, \emph{Multipole Theory in Electromagnetism:
  Classical, Quantum, and Symmetry Aspects, with Applications}, ser. Oxford
  Science Publications.\hskip 1em plus 0.5em minus 0.4em\relax {Oxford : New
  York}: {Clarendon Press ; Oxford University Press}, 2005, no. 128.

\bibitem{agranovich2013crystal}
V.~M. Agranovich and V.~Ginzburg, \emph{Crystal optics with spatial dispersion,
  and excitons}.\hskip 1em plus 0.5em minus 0.4em\relax Springer Science \&
  Business Media, 2013, vol.~42.

\bibitem{simovski2018composite}
C.~Simovski, \emph{Composite Media with Weak Spatial Dispersion}.\hskip 1em
  plus 0.5em minus 0.4em\relax CRC Press, 2018.

\bibitem{choContributionElectricQuadrupole2008}
D.~J. Cho, F.~Wang, X.~Zhang, and Y.~R. Shen, ``Contribution of the electric
  quadrupole resonance in optical metamaterials,'' \emph{Phys. Rev. B},
  vol.~78, no.~12, p. 121101, Sep. 2008.

\bibitem{petschulatUnderstandingElectricMagnetic2010}
J.~Petschulat, J.~Yang, C.~Menzel, C.~Rockstuhl, A.~Chipouline, P.~Lalanne,
  A.~T{\"u}ennermann, F.~Lederer, and T.~Pertsch, ``Understanding the electric
  and magnetic response of isolated metaatoms by means of a multipolar field
  decomposition,'' \emph{Optics Express}, vol.~18, no.~14, p. 14454, Jul. 2010.

\bibitem{muhligMultipoleAnalysisMetaatoms2011}
S.~M{\"u}hlig, C.~Menzel, C.~Rockstuhl, and F.~Lederer, ``Multipole analysis of
  meta-atoms,'' \emph{Metamaterials}, vol.~5, no. 2-3, pp. 64--73, Jun. 2011.

\bibitem{grahnElectromagneticMultipoleTheory2012}
P.~Grahn, A.~Shevchenko, and M.~Kaivola, ``Electromagnetic multipole theory for
  optical nanomaterials,'' \emph{New Journal of Physics}, vol.~14, no.~9, p.
  093033, Sep. 2012.

\bibitem{evlyukhinCollectiveResonancesMetal2012}
A.~B. Evlyukhin, C.~Reinhardt, U.~Zywietz, and B.~N. Chichkov, ``Collective
  resonances in metal nanoparticle arrays with dipole-quadrupole
  interactions,'' \emph{Phys. Rev. B}, vol.~85, no.~24, p. 245411, Jun. 2012.

\bibitem{yaghjianHomogenizationSpatiallyDispersive2013}
A.~D. Yaghjian, A.~Al{\`u}, and M.~G. Silveirinha, ``Homogenization of
  spatially dispersive metamaterial arrays in terms of generalized electric and
  magnetic polarizations,'' \emph{Photonics and Nanostructures - Fundamentals
  and Applications}, vol.~11, no.~4, pp. 374--396, Nov. 2013.

\bibitem{silveirinhaBoundaryConditionsQuadrupolar2014}
M.~G. Silveirinha, ``Boundary conditions for quadrupolar metamaterials,''
  \emph{New Journal of Physics}, vol.~16, no.~8, p. 083042, Aug. 2014.

\bibitem{Koendrink2014Retrieval}
F.~Bernal~Arango, T.~Coenen, and A.~F. Koenderink, ``Underpinning hybridization
  intuition for complex nanoantennas by magnetoelectric quadrupolar
  polarizability retrieval,'' \emph{ACS Photonics}, vol.~1, no.~5, pp.
  444--453, 2014.

\bibitem{poutrina2014multipole}
E.~Poutrina and A.~Urbas, ``Multipole analysis of unidirectional light
  scattering from plasmonic dimers,'' \emph{Journal of Optics}, vol.~16,
  no.~11, p. 114005, 2014.

\bibitem{babichevaMetasurfacesElectricQuadrupole2018}
V.~E. Babicheva and A.~B. Evlyukhin, ``Metasurfaces with {{Electric
  Quadrupole}} and {{Magnetic Dipole Resonant Coupling}},'' \emph{ACS
  Photonics}, vol.~5, no.~5, pp. 2022--2033, May 2018.

\bibitem{achouri2021extension}
K.~Achouri and O.~J. Martin, ``Extension of lorentz reciprocity and poynting
  theorems for spatially dispersive media with quadrupolar responses,''
  \emph{arXiv preprint arXiv:2102.08197}, 2021.

\bibitem{holloway2012overview}
C.~Holloway, E.~F. Kuester, J.~Gordon, J.~O'Hara, J.~Booth, and D.~Smith, ``An
  overview of the theory and applications of metasurfaces: the two-dimensional
  equivalents of metamaterials,'' \emph{IEEE Antennas Propag. Mag.}, vol.~54,
  no.~2, pp. 10--35, April 2012.

\bibitem{Pfeiffer2013a}
C.~Pfeiffer and A.~Grbic, ``Metamaterial {H}uygens' surfaces: tailoring wave
  fronts with reflectionless sheets,'' \emph{Phys. Rev. Lett.}, vol. 110, p.
  197401, May 2013.

\bibitem{PhysRevApplied.2.044011}
C.~Pfeiffer and A.~Grbic, ``Bianisotropic metasurfaces for optimal polarization
  control: Analysis and synthesis,'' \emph{Phys. Rev. Applied}, vol.~2, p.
  044011, Oct 2014.

\bibitem{6905746}
M.~Selvanayagam and G.~Eleftheriades, ``Polarization control using tensor
  {H}uygens surfaces,'' \emph{IEEE Trans. Antennas Propag.}, vol.~62, no.~12,
  pp. 6155--6168, Dec 2014.

\bibitem{Achouri2015c}
K.~Achouri, B.~A. Khan, S.~Gupta, G.~Lavigne, M.~A. Salem, and C.~Caloz,
  ``Synthesis of electromagnetic metasurfaces: principles and illustrations,''
  \emph{EPJ Applied Metamaterials}, vol.~2, p.~12, 2015.

\bibitem{achouri2018design}
K.~Achouri and C.~Caloz, ``Design, concepts, and applications of
  electromagnetic metasurfaces,'' \emph{Nanophotonics}, vol.~7, no.~6, pp.
  1095--1116, 2018.

\bibitem{Idemen1973}
M.~M. Idemen, \emph{Discontinuities in the Electromagnetic Field}.\hskip 1em
  plus 0.5em minus 0.4em\relax John Wiley \& Sons, 2011.

\bibitem{kuester2003av}
E.~F. Kuester, M.~Mohamed, M.~Piket-May, and C.~Holloway, ``Averaged transition
  conditions for electromagnetic fields at a metafilm,'' \emph{IEEE Trans.
  Antennas Propag.}, vol.~51, no.~10, pp. 2641--2651, Oct 2003.

\bibitem{achouri2020fundamental}
K.~Achouri and O.~J. Martin, ``Fundamental properties and classification of
  polarization converting bianisotropic metasurfaces,'' \emph{arXiv preprint
  arXiv:2008.05776}, 2020.

\bibitem{savinov_toroidal_2014}
V.~Savinov, V.~A. Fedotov, and N.~I. Zheludev, ``Toroidal dipolar excitation
  and macroscopic electromagnetic properties of metamaterials,'' \emph{Phys.
  Rev. B}, vol.~89, no.~20, p. 205112, May 2014.

\bibitem{achouri2015improvement}
K.~Achouri, M.~A. Salem, and C.~Caloz, ``Improvement of metasurface continuity
  conditions,'' in \emph{2015 International Symposium on Antennas and
  Propagation (ISAP)}.\hskip 1em plus 0.5em minus 0.4em\relax IEEE, 2015, pp.
  1--3.

\bibitem{albooyeh2016electromagnetic}
M.~Albooyeh, S.~Tretyakov, and C.~Simovski, ``Electromagnetic characterization
  of bianisotropic metasurfaces on refractive substrates: General theoretical
  framework,'' \emph{Ann. Phys.}, vol. 528, no. 9-10, pp. 721--737, 2016.

\bibitem{jackson_classical_1999}
J.~D. Jackson, \emph{Classical electrodynamics}, 3rd~ed.\hskip 1em plus 0.5em
  minus 0.4em\relax New York, {NY}: Wiley, 1999.

\bibitem{papas2014theory}
C.~H. Papas, \emph{Theory of electromagnetic wave propagation}.\hskip 1em plus
  0.5em minus 0.4em\relax Courier Corporation, 2014.

\bibitem{serdkov2001electromagnetics}
A.~Serd\t{iu}kov, I.~Semchenko, S.~Tretyakov, and A.~Sihvola,
  \emph{Electromagnetics of bi-anisotropic materials-Theory and
  Application}.\hskip 1em plus 0.5em minus 0.4em\relax Gordon and Breach
  science publishers, 2001, vol.~11.

\bibitem{arfken1999mathematical}
G.~B. Arfken and H.~J. Weber, ``Mathematical methods for physicists,'' 1999.

\bibitem{kong1986electromagnetic}
J.~Kong, \emph{Electromagnetic wave theory}, ser. A Wiley-Interscience
  publication.\hskip 1em plus 0.5em minus 0.4em\relax John Wiley \& Sons, 1986.

\bibitem{Jackson1998}
J.~D. Jackson, \emph{Classical Electrodynamics}, 3rd~ed.\hskip 1em plus 0.5em
  minus 0.4em\relax Wiley, 1998.

\bibitem{calozElectromagneticNonreciprocity2018}
C.~Caloz, A.~Al{\`u}, S.~Tretyakov, D.~Sounas, K.~Achouri, and Z.-L.
  {Deck-L{\'e}ger}, ``Electromagnetic {{Nonreciprocity}},'' \emph{Physical
  Review Applied}, vol.~10, no.~4, Oct. 2018.

\bibitem{LightPropagationCubic1990}
``Light propagation in cubic and other anisotropic crystals,''
  \emph{Proceedings of the Royal Society of London. Series A: Mathematical and
  Physical Sciences}, vol. 430, no. 1880, pp. 593--614, Sep. 1990.

\bibitem{rothwell2018electromagnetics}
E.~J. Rothwell and M.~J. Cloud, \emph{Electromagnetics}.\hskip 1em plus 0.5em
  minus 0.4em\relax CRC press, 2018.

\end{thebibliography}

\end{document}